\documentclass[usegraphicx,useAMS,usenatbib]{mn2e}

\usepackage{graphicx}
 
\title[Is there period-doubling in Blazhko stars?]{Is the apparent period-doubling in Blazhko stars actually an illusion?}

\author[Paul H. Bryant]
{Paul H. Bryant\thanks{E-mail: pbryant@ucsd.edu}\\
BioCircuits Institute (formerly Institute for Nonlinear Science), University of California, San Diego, La Jolla, CA 92093, USA}

\begin{document}

\date{Accepted \_\_\_\_\_\_\_\_\_\_\_\_\_\_. Received \_\_\_\_\_\_\_\_\_\_\_\_\_\_; in original form \_\_\_\_\_\_\_\_\_\_\_\_\_\_}

\pagerange{\pageref{firstpage}--\pageref{lastpage}} \pubyear{2015}
  
\maketitle

\label{firstpage}
 
\begin{abstract}
The light curves of many Blazhko stars exhibit intervals in which successive pulsation maxima alternate between two levels in a way that is characteristic of period-doubling. In addition, hydrocode models of these stars have clearly demonstrated period-doubling bifurcations. As a result, it is now generally accepted that these stars do indeed exhibit period-doubling. Here we present strong evidence that this assumption is incorrect. The alternating maxima likely result from the presence of one or more near-resonant modes which appear in the stellar spectra and are slightly but significantly offset from 3/2 times the fundamental frequency.  We show that a previously proposed explanation for the presence of these peaks is inadequate. The phase-slip of the dominant near-resonant peak in RR Lyr is shown to be fully correlated with the parity of the observed alternations, providing further strong evidence that the process is nonresonant and cannot be characterized as period-doubling.  The dominant near-resonant peak in V808 Cyg has side-peaks spaced at twice the Blazhko frequency.  This apparent modulation indicates that the peak corresponds to a vibrational mode and also adds strong support to the beating-modes model of the Blazhko effect which can account for the doubled frequency.  The modulation also demonstrates the ``environment" altering effect of large amplitude modes which is shown to be consistent with the amplitude equation formalism. 
\end{abstract}


\section{Introduction}
\label{introduction}
It is well known that some nonlinear dynamical systems can exhibit a period-doubling bifurcation \citep[see, e.g.][]{Strogatz}.  For parameter values below the bifurcation point, the system is oscillating periodically at frequency $f$ and may also have spectral peaks at the harmonics of this frequency.  In passing the bifurcation point, a small peak at $f/2$ emerges and grows in size.  The oscillations exhibited by the system alternate between two slightly different paths in the phase space, so that the period has suddenly become twice as long as it was originally.  Typically this is observed in the successive maxima of the oscillation which alternate between two slightly different levels.  Odd harmonics of $f/2$ typically appear in the spectrum as well.  We will refer to these as half-integer frequencies.  Often additional period-doublings occur as the parameter is further advanced resulting in the appearance of $f/4$, $f/8$, etc.  This cascade typically goes all the way to $f/\infty$ over a finite change of the parameter and beyond this point the dynamics do not repeat and the system is called ``chaotic".

\citet{Kolenberg2010} first reported seeing alternating maxima in the light curves of some Blazhko stars and \citet{Szabo2010} demonstrated period-doubling in a hydrocode model of RR Lyr.  Since then it has been generally accepted that the observed alternation effect was indeed caused by a period-doubling bifurcation \citep[see, e.g.,][]{Buchler2011, Kollath2011, Guggenberger2012, Molnar2012, Kolenberg2012, Benko, Szabo2014, Leborgne}.  In this paper we show the serious problems with this claim, and present strong evidence to support an alternate explanation, namely that the alternation is caused by the presence of one or more excited modes whose frequencies are close to 3/2 times the fundamental frequency and (in most cases at least) are not in resonance with the fundamental mode.  The largest of these peaks have frequencies that can be accurately determined and can be observed over the entire \emph{Kepler} data set of about four years.  Other studies have noticed that these peaks are off resonance, but a seemingly plausible explanation has been offered \citep[][Section 3.2]{Szabo2010} and as a result the idea persists that presence of these near-resonant peaks indicates that a period-doubling bifurcation has occurred.
This belief is strengthened by the fact that actual period-doubling has been observed in hydrocode models \citep{Szabo2010, Kollath2011, Smolec}, and it is assumed that therefore the same must be occurring in actual RR Lyrae stars.
But as is shown below, these hydrocode results deviate quite substantially from the results obtained from actual stellar data and thus they do not prove that period-doubling is occurring in the real stars.

The argument by \citet{Szabo2010} is that the half-integer frequencies appear to be off-resonance because they are effectively modulated, periodically and/or randomly, and that this produces a cluster of side-peaks in the spectrum.  It does not explain however, how the central peak can be shifted to one side of the exact half-integer frequency.  Even for a completely random modulation the result should be a cluster of peaks that is centered on the base frequency.  This is in fact what happens in their simulation of the effect as seen in their Figure 10, where the cluster of peaks in part (b) does line up with the main peak in part (a), and if the image is magnified it can be seen that the main peak in part (a) is in precise alignment with one of the peaks in part (b), and therefore this peak is exactly on-resonance.  (Of the two tallest peaks in the cluster, it is the one on the left.)

We will show that the dominant peak near $(3/2) f_0$ in RR Lyr is substantially off-resonance and persists over the entire Kepler data set and is therefore not an artifact of some random or chaotic process.  We will also show that the observed half-integer oscillation is not phase locked to the fundamental but is constantly slipping in phase at the rate predicted by the offset of the peak location from exact resonance.  We also show that a similar off-resonance peak exists in the spectrum for V808 Cyg, another star which is believed to exhibit a strong period-doubling effect \citep{Szabo2010}.  This peak has side-peaks consistent with \emph{modulation at double the Blazhko frequency}.  One explanation for his strange doubling effect (described in detail below) is that this peak does in fact correspond to a nonresonant mode and that the side peaks result from interactions between this mode and the modes that generate the Blazhko effect as described by the ``beating-modes model" \citep{Bryant2015}.\footnote{Not to be confused with \citet{Bryant2014}, an earlier work by the author on the Blazhko effect, which \citet{Bryant2015} reinterprets and supersedes.}  It thus simultaneously presents evidence against period-doubling and evidence in favor of the beating-modes Blazhko model (since it would appear that other models cannot provide an explanation for the doubled frequency).  The analysis introduces the concept of the ``environment" within which the modulated mode resides and which is generated by the larger amplitude Blazhko modes.  We show that this concept is compatible with the amplitude equation formalism \citep{Buchler1984}.

We note that the presence of a near-resonant mode in the beating-modes model is very similar to the currently proposed explanation for the maxima alternation, but aside from this similarity the two explanations are not linked,  i.e. the validity of the current work does not depend on that of the Blazhko explanation or vice-versa.  Note that in the beating-modes model there can be nonlinear interactions between these two modes as long as these interactions do not cause phase-locking.  Some nonlinear interactions will depend on the relative phase of the two modes, but since this is constantly changing, it will typically have a modulatory effect, going through one modulation cycle with the periodicity of the relative phase.  But this modulation is a secondary effect, the primary cause of the Blazhko effect is the phase-slip itself.  \citet{Bryant2015} reproduced with good accuracy the Blazhko effect of RR Lyr without including any actual modulation.  One might assume then that such modulation effects are relatively small for RR Lyr, but this is not necessarily the case for other Blazhko stars.

\section{Results and Discussion}
\label{results}
\begin{figure}
\begin{center}
\includegraphics[width=1.00\columnwidth]{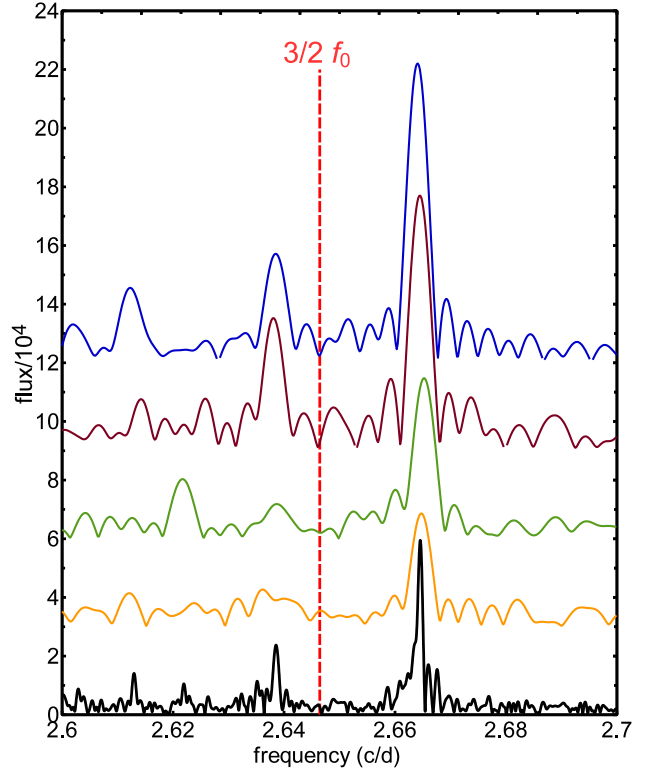}
\caption{\label{fig1}Spectra for RR Lyr in the vicinity of $(3/2) f_0$ computed from \emph{Kepler} project KIC 7198959, long cadence corrected flux data using Period04 software.  Average (or zero point) flux for this data is about $1.2222 \times 10^7$.  Data was pre-whitened, removing the fundamental and its harmonics through tenth.  The upper curve is computed from quarters Q5, Q6 and Q7; it is offset by 12 units vertically for clarity.  The second curve uses Q7, Q8 and Q9 with offset 9.  The third curve uses Q11, Q12 and Q13 with offset 6.  The fourth curve uses Q14, Q15 and Q16 with offset 3.  The fifth curve uses all available data: Q1 through Q17 with no offset.  Note that Q1 and Q17 are partial quarters and Q3, Q4 and Q10 are not available.  The dashed vertical line marks the location of $(3/2) f_0$.  Note that none of the five curves have a peak at this location, while they all have one at about 2.6645 which is clearly nonresonant.}
\end{center}
\end{figure}

In Figure~\ref{fig1} we show spectra for RR Lyr.
The first four curves from the top down were calculated for groups of three consecutive quarters of \emph{Kepler} data as indicated in the caption.  The bottom curve was calculated using all available Kepler data, quarters 1 through 17, and thus has much higher frequency resolution.  Note the consistent pattern in this set of curves with a dominant peak at 2.6645, henceforth called $f_A$, while the expected location for resonance would be at 2.6464 (marked with a dashed vertical line in the figure).  We propose that $f_A$ is the base frequency of a mode which we will refer to as Mode A.  The frequency difference being 0.0181 means that this oscillation will slip in phase by one half cycle about every 27.6 days.  This phase slip can be observed in the alternation pattern, since a half cycle slip will cause a switch from the even cycles being high to the odd cycles being high.  \citet{Leborgne} also found this frequency (0.018) in their spectral analysis of the alternation effect (see their Figure 3).  We verify this phase slip effect in Table~\ref{tab1}, which shows a full correlation in the data for RR Lyr between the high cycles being odd or even and the number of half cycle slips being odd or even.
\begin{table}
  \caption{Correlation between the phase slip of the near resonant mode and the phase of the maxima alternation pattern for RR Lyr.  A single point is chosen near the middle of each low-high alternation sequence.  It is required to be the point of maximum flux in a ``high" cycle.  In some cases the sequence may include a few bad or non-alternating points provided the alternation continues with the proper phase.  The first point in the table has truncated Barycentric Julian Date 54992.569.  Columns: time is the time in days relative to the first point; len is the approximate length in cycles of the alternation sequence;  cyc is the cycle number, calculated by multiplying time by $f_0$ (1.76429291); par1 is the parity of cyc after rounding to the nearest integer; slip is the phase slip in half cycles, calculated by multiplying time by twice the frequency offset of the near resonant mode ($2 \times 0.0181$) and adding a correction of -0.3 (chosen to make the best overall fit to the entire set); par2 is the parity of slip after rounding to the nearest integer; cor is the correlation of par1 and par2, indicated as ``yes" if they match and ``no" if they do not.}
  \begin{tabular}{lllllll}
\hline
time&len&cyc&par1&slip&par2&cor \\ \hline
0&12&0&even&-0.3&even&yes\\
36.843&24&65.001&odd&1.034&odd&yes\\
65.715&16&115.941&even&2.079&even&yes\\
285.658&12&503.985&even&10.041&even&yes\\
306.644&54&541.011&odd&10.801&odd&yes\\
338.931&24&597.973&even&11.969&even&yes\\
392.244&32&692.033&even&13.899&even&yes\\
424.55&30&749.03&odd&15.069&odd&yes\\
456.814&32&805.954&even&16.237&even&yes\\
540.159&18&953&odd&19.254&odd&yes\\
587.789&11&1037.032&odd&20.978&odd&yes\\
618.95&35&1092.009&even&22.106&even&yes\\
653.504&26&1152.972&odd&23.357&odd&yes\\
669.974&17&1182.03&even&23.953&even&yes\\
701.136&20&1237.01&odd&25.081&odd&yes\\
731.175&43&1290.007&even&26.169&even&yes\\
926.722&14&1635.009&odd&33.247&odd&yes\\
1036.675&20&1828.998&odd&37.228&odd&yes\\ \hline
  \end{tabular}
  \label{tab1}
\end{table}
Most, if not all, of the easily visible alternation sequences of length longer than 10 are included in the table.  Note that this continuous phase slip means that there is no phase-locking between the fundamental and Mode A and \emph{therefore there is no resonance and no period-doubling}.  The fact that this effect can be followed through the entire 17 quarters of \emph{Kepler} data indicates that this is tied to something with a very stable frequency (i.e. a non-resonant mode).  It can't be attributed to a random phase drift and it seems quite incompatible with period doubling.

Note that the alternation amplitude goes to zero at the transition point between where the high cycles are even and where they are odd.  This happens again when they transition back to even, and so on.  The amplitude of Mode A has not gone to zero at these points, but its relative phase is such that it is not visible as a maxima alternation.

An interesting thing to note in Figure~\ref{fig1} is that the peak $f_A$ is diminishing in height over time.  The amplitude of the Blazhko effect is also decreasing simultaneously.  This might be taken as evidence of a possible connection between them, but it could also be just a coincidence.  \citet{Bryant2015} has shown that the Blazhko effect may also be an artifact of a near-resonant nonradial mode.  So one simple explanation for the correlation could be that the excitation levels of all of these nonradial modes are decreasing for some reason.  For example, if the magnetic field of the star was slowly increasing in time, this might cause a slow increase in turbulence resulting in increased damping and/or decreased excitation of the modes and this would affect both the Blazhko effect and the maxima alternation.  A similar magnetic mechanism was recently proposed by \citet{Stothers2006, Stothers2010} as part of an explanation of the Blazhko effect.  One critique of this explanation \citep{Molnar2012} claimed that the mechanism was too slow to explain the Blazhko effect, but this would not disqualify it from explaining these long term fluctuations in the Blazhko effect.  But whatever the reason, long term variations in the Blazhko effect in RR Lyr have been known for some time and remain unexplained; see e.g. \citet{Leborgne}.  Period doubling does not provide a definitive explanation for this mystery either.  The observed correlation between the diminishing of $f_A$ and the diminishing of the Blazhko effect could be real or it could be a coincidence, there is insufficient information to decide based on data coming from just one star.  However, if it was real, it could explain why the maxima alternations have thus far only been found in Blazhko stars.

\begin{figure}
\begin{center}
\includegraphics[width=1.00\columnwidth]{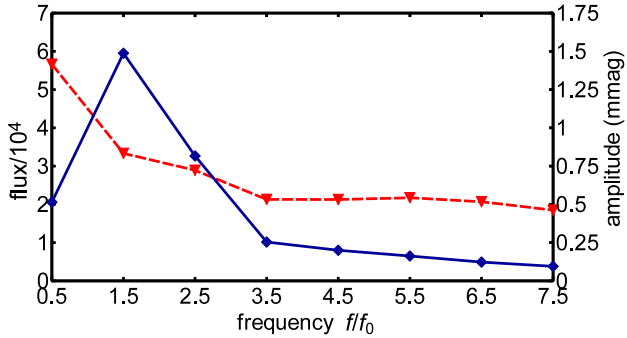}
\caption{\label{fig2}Comparison of the observed half-integer frequency peaks with those of a hydrocode model exhibiting period-doubling.  The solid curve with diamond data points is calculated from \emph{Kepler} data for RR Lyr using all available quarters, Q1 through Q17.  The vertical scale for this curve is on the left.  These peaks are off-resonance as discussed in the text.  The dashed curve with inverted triangle data points is replotted from the M8 curve in Figure 11 in \citet{Smolec}.  The vertical scale for this curve is on the right.  These peaks are exactly on-resonance.  Note the great disparity between the two curves, the first having a strong maxima at 3/2 while the second has a strong maxima at 1/2.}
\end{center}
\end{figure}

The spectrum also shows additional peaks at frequencies that are mixing products of this peak with the fundamental and its harmonics, i.e. frequencies of the form $f_A + nf_0$ where $n$ is an integer and $f_0$ is the fundamental frequency.  These peaks are all near, but not equal to, the half-integer frequencies that are expected for period-doubling.  The amplitudes of these peaks are determined, based on all available data, and the results compared in Figure~\ref{fig2} with the corresponding spectrum for period-doubling in a hydrocode model of RR Lyr \citep[taken from M8 in Figure 11 in][]{Smolec}.
Note that their spectrum has the dominant peak at $(1/2) f_0$ rather than at $(3/2) f_0$.  The ratio of the peak heights (the 3/2 peak height over the 1/2 peak height) changes rather drastically between the two cases. This lack of agreement in the spectrum can be taken as fairly strong evidence that the model is not presenting a correct representation of the dynamics of the actual star.  Having the maximum half-integer peak near 3/2 in the observational spectrum is, of course, consistent with the source for this peak being an excited mode with that frequency.  The other half-integer peaks are then nonlinear mixing peaks, which would typically be expected to be smaller in amplitude than the one near 3/2 in agreement with observations. Other observational results also have the maximum at 3/2 rather than 1/2 \citep[][Figure 8]{Szabo2010}.  The \citet{Smolec} results also show an exact correlation between the Blazhko phase of their model and the appearance and disappearance of the period-doubling, something that is entirely missing in the actual stellar data.  This is of course only relevant if we assume that the Blazhko effect involves actual modulation as opposed to a beating-modes process \citep{Bryant2015}.

As was noted by \citet{Szabo2010}, when observed on a short timescale the amplitudes of the half-integer frequency peaks appear to be modulated in an apparently random fashion (see their Figure 3, bottom panel).  Here are three possible reasons (all of which could be acting simultaneously):  1) Turbulence noise is interacting with Mode A and causing these fluctuations.  Since this mode is only marginally unstable, it will be very sensitive and can effectively amplify this noise.  \citet{Buchler1993} previously studied such stochastic excitation of stellar pulsations.  2) Some of the nearby peaks in the spectrum may correspond to separate excited modes and add additional near-resonant frequencies into the system.  Several of these added together could generate something that looks like a random modulation.  3) Noise and measurement error of the \emph{Kepler} electronics could be affecting the results.

This modulation or apparent modulation of Mode A also means that occasionally the observed maxima alternations are considerably larger than might be expected from its peak height in the spectrum.  But since the time series can \emph{always} be reconstructed from the spectral information, these occasional large alternations must be reflected in the combination of the main peak with various associated side peaks and harmonics when these are all in phase.

An interesting question is why $f_A$ is so close in frequency to $(3/2) f_0$.  The usual mechanism for transferring energy from the fundamental mode to another mode only works when there is phase-locking between the modes and that is apparently not happening in this case.  But perhaps there is some other, as yet unknown, mechanism at work.  It is also possible that the proximity is merely an accident.  In Figure~\ref{fig1}, note the two smaller peaks to the left of the main peak.  There are more peaks further to the left (off scale) of a similar size and more densely packed together.  This suggests that there is a cluster of peaks that just happens to slightly overlap the resonance point.  Other stars, most notably V445 Lyr, appear to have lots of excited modes that are far from resonance; see e.g. \citet{Guggenberger2012}.  If a mode can be excited in arbitrary locations then near-resonant locations should also be possible.

\begin{figure}
\begin{center}
\includegraphics[width=1.00\columnwidth]{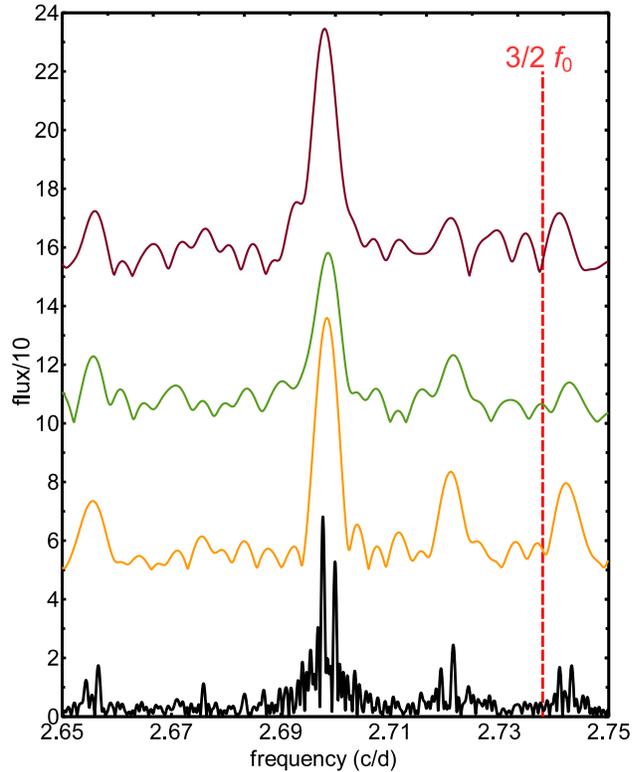}
\caption{\label{fig3}Spectra for V808 Cyg in the vicinity of $(3/2) f_0$ computed from \emph{Kepler} project KIC 4484128, long cadence corrected flux data using Period04 software.  Average (or zero point) flux for this data is about $8.8363 \times 10^3$.  Data was pre-whitened, removing the fundamental and its harmonics through tenth.  The upper curve is computed from quarters Q2, Q3 and Q4; it is offset by 15 units vertically for clarity.  The second curve uses Q7, Q8 and Q9 with offset 10.  The third curve uses Q11, Q12 and Q13 with offset 5.  The fourth curve uses all available data: Q1 through Q17 with no offset.  Note that Q1 and Q17 are partial quarters and Q6, Q10 and Q14 are not available.  The dashed vertical line marks the location of $(3/2) f_0$.  Note that none of the five curves have a peak at this location, while they all have one at about 2.6983 which is clearly nonresonant.  See text for discussion of the split peak in the bottom curve.}
\end{center}
\end{figure}

In Figure~\ref{fig3} we show spectra for V808 Cyg.
The first three curves from the top down were calculated for groups of three consecutive quarters of \emph{Kepler} data as indicated in the caption.  The bottom curve was calculated using all available Kepler data, and thus has much higher frequency resolution.  Note the consistent pattern in this set of curves, with a dominant peak at 2.6983, which we will again call $f_A$, while the expected location for resonance would be at 2.7379 (marked with a dashed vertical line in the figure).  This time the dominant peak is found on the opposite side of the resonance compared to Figure~\ref{fig1}.  An interesting feature of the bottom curve is that the $f_A$ has split into two peaks with frequencies of $2.6977$ and $2.6999$.  One explanation would be that the noisy turbulent environment of the star is causing the amplitude to fluctuate in a way that happens to look like the sum of these two frequencies.  Another explanation is that there really are two closely spaced modes that generate these peaks.  In favor of the second explanation is the fact that the difference between the peak frequencies is 0.0022 and from this one can determine that the two modes slipped in phase by about 3.2 cycles over the time span of the full data set.  This seems like an excessive number of consistent cycles to be accidentally generated by a random process.  A second interesting feature is the fact that the pair of peaks appears to have pairs of side peaks.  A total of 8 side peaks (four pairs) are visible (although one is barely visible above the noise).  The fact that it appears modulated is a strong indication that this peak (or pair of peaks) does correspond to an actual vibrational mode (or pair of modes) as opposed to the explanation that it is some kind of noisy period-doubling byproduct.

Perhaps even more interesting is that the apparent modulation frequency, based on the side-peak spacing, is twice the Blazhko frequency.  By our measurements from the entire data set, the Blazhko frequency is about 0.01085, so twice this is 0.02170.  We measure the side-peak spacing to be about 0.02165 so this is a very accurate match.  We have an explanation for this doubled frequency in terms of the beating-modes model of the Blazhko effect \citep{Bryant2015}.  That model proposes that the apparent modulation of the fundamental is actually the beat frequency produced by simply adding the fundamental mode oscillation to that of a near-resonant (presumably nonradial) mode, whose frequency differs from the fundamental by the Blazhko frequency.  As we explain below, the modulation of Mode A observed in the current problem may be due to the type of nonlinear interaction that is not phase-dependent.

Let us assume, to be specific, that the nonradial mode is a dipolar mode with axial symmetry, i.e. a mode with indices $l=1$ and $m=0$ \citep[see, e.g.][Section 4]{Unno}.  The dipolar character will cause the radial component of the oscillations in the northern hemisphere to be be 180 degrees out of phase with those in the southern hemisphere.  If, at one point in time, this oscillation has a particular phase relationship with the fundamental mode in the northern hemisphere, then one half of a Blazhko period later it have the identical phase relationship with the fundamental mode in the southern hemisphere.  So for Mode A, these two points in time present a spatially inverted but otherwise equivalent ``environment" for that mode to reside in.  Thus if there is any variation in this environment then it must repeat every half Blazhko cycle, i.e. it must oscillate at twice the Blazhko frequency.

This type of oscillatory environment can have a modulatory effect on the frequency of Mode A.  This can be seen through the cross coupling ``$T$" terms that appear in the amplitude equations, see, e.g. Equation (1) \citet{Buchler2011}, Equations (1a) and (1b) in \citet{Moskalik1990} and Equations (70) and (71) in \citet{Buchler1984}.  In the equation for $da/dt$, this term has the form $T|b|^2a$ where $T$ is a complex coefficient and $a$ and $b$ are the complex amplitudes of Modes A and B.  The imaginary component of $T$ results in the frequency of Mode A being shifted in proportion to the square of the amplitude of Mode B, while the real component of $T$ results in a similar shift in the growth rate of Mode A.  Note that this term, unlike some coupling terms, is a non-resonant term.  By this we mean that it does not depend on the relative phase of Mode A and will not cause phase-locking between the modes.  This is in contrast, for example, to the ``C" terms appearing in Equation (1) \citet{Buchler2011} which can result in phase-locking between modes which might otherwise be slightly off resonance.    So the presence of the $T$ term in the equation for $da/dt$ can be thought of as an effect of the ``environment" generated by Mode B on the dynamics of Mode A.

Extending this analysis to the current problem with three modes, we will let Mode B be the fundamental mode, and let Mode C be the mode that is near-resonant with the fundamental (according to the beating-modes model), while Mode A retains its definition from earlier in the paper.  To represent a combined effect of modes B and C in generating an environment for mode A, we may add $(U_1b^*c+U_2c^*b)a$ to the equation for $da/dt$, where $U_1$ and $U_2$ are complex coefficients.  Unlike the $T$ terms, these generate an environment that is slowly varying and will therefore slowly change the frequency and/or amplitude of Mode A.  Since the frequency difference between Modes B and C is the Blazhko frequency, the result will be to modulate Mode A at that frequency.  If we assume that mode C is a dipolar mode, then one might expect, as a result of the environmental symmetry arguments given above, that the coefficients $U_1$ and $U_2$ for this case would be exactly zero.  However these arguments would not apply for axisymmetric modes with even symmetry about the equator, i.e. those with even degree $l$ and order $m=0$.  So for those cases these terms could produce modulation of mode A at the Blazhko frequency.  For axisymmetric modes of odd degree $l$ (including the dipolar modes) the modulation frequency should be twice the Blazhko frequency and the lowest order environmental terms that could produce this result would appear to be of the form $(V_1b^{*2}c^2$ +$V_2c^{*2}b^2)a$.  Note that as the degree $l$ increases, this distinction between odd and even diminishes in importance and so the $U$ coefficients for even $l$ should tend toward zero.  Also note that modes with nonzero order $m$ cannot generate this type of oscillatory environment in conjunction with the fundamental unless they are excited in pairs with whose $m$ values add to zero.  In this case there would again be a symmetry that would only allow modulation at twice the Blazhko frequency.


An interesting question is whether the symmetry properties of the dynamics are retained when the amplitude is large and therefore strongly nonlinear.  This may be indicated in the current case by the highly non-sinusoidal nature of the fundamental oscillation.  Studies of simpler nonlinear systems have indicated that symmetrical quasiperiodic oscillations may be quite common; see e.g. Figure (2b) in \citet{Bryant1987}.  As discussed in \citet{Bryant1984, Bryant1987} another possibility is complementary asymmetric attractor pairs, though this would seem unlikely except when phase locking occurs; see e.g. Figure (3b) in \citet{Bryant1987}.

\label{lastpage}
 
\end{document}